# Towards local electromechanical probing of cellular and biomolecular systems in a liquid environment

Sergei V. Kalinin,[*] Brian J. Rodriguez, Stephen Jesse, and Katyayani Seal

Materials Sciences and Technology Division and the Center for Nanophase Materials Sciences, Oak Ridge National Laboratory, Oak Ridge, TN 37931

Roger Proksch, Sophia Hohlbauch, and Irene Revenko

Asylum Research, Santa Barbara, CA 93117

Gary Lee Thompson and Alexey A. Vertegel

Clemson University, Department of Bioengineering, Clemson, SC 29634

Electromechanical coupling is ubiquitous in biological systems with examples ranging from simple piezoelectricity in calcified and connective tissues to voltage-gated ion channels, energy storage in mitochondria, and electromechanical activity in cardiac myocytes and outer hair cell stereocilia. Piezoresponse force microscopy (PFM) has originally emerged as a technique to study electromechanical phenomena in ferroelectric materials, and in recent years, has been employed to study a broad range of non-ferroelectric polar materials, including piezoelectric biomaterials. At the same time, the technique has been extended from ambient to liquid imaging on model ferroelectric systems. Here, we present results on local electromechanical probing of several model cellular and biomolecular systems, including insulin and lysozyme amyloid fibrils, breast adenocarcinoma cells, and bacteriorhodopsin in a liquid environment. The specific features of SPM operation in liquid are delineated and bottlenecks on the route towards nanometer-resolution electromechanical imaging of biological systems are identified.

---

[*] Corresponding author, sergei2@ornl.gov

# I. Introduction

Coupling between electrical and mechanical phenomena is ubiquitous in biological systems. Multiple examples include phenomena such as voltage controlled muscular contractions [1], cell electromotility [2], and electromotor proteins [3]. In many cases, these couplings are the functional bases for processes such as transduction of acoustic signals into electrical pulses in the outer hair cell stereocilia, cardiac activity, energy storage in mitochondria, etc. On a more fundamental level, cellular membranes are flexoelectric, i.e. possess linear coupling between electric field and strain gradient (or local radius of curvature) [4]. Biopolymers forming calcified and connective tissues are often piezoelectric [5], i.e. exhibit linear coupling between electric field and strain. Significant effort has been directed towards understanding origins of electromechanical activity of these types of biological systems and its relationship to functionality [6-15]. However, due to the complex spatial structure of biological systems, this task requires real-space probing of electromechanical coupling on the nanoscale.

Piezoresponse Force Microscopy (PFM) has emerged in the last decade as a primary tool for characterization of static and dynamic polarization phenomena in ferroelectric materials based on local electromechanical coupling.[16-23] PFM employs a conducting tip to probe the mechanical response of a surface to an applied bias. In ferroelectric materials, electromechanical response is directly coupled to the primary polarization order parameter through electrostrictive constants. Hence, mapping of local electromechanical activity and its evolution under applied bias (Piezoresponse spectroscopy) and force provides insight into local polarization dynamics [24-30]. In the last several years, PFM was extended to a number of polar piezoelectric materials such as III-V nitrides [31], ferroelectric polymers [32], and piezoelectric biopolymers [33-35]. In these systems, electromechanical activity is strongly related to local surface termination (III-V nitrides), presence and molecular orientation of electromechanically active proteins, etc. Probing the electromechanical activity can in turn provide an insight into nanoscale surface structure and functionality.

One of the key challenges in PFM imaging of biological systems is imaging in liquid environment. This is required both to maintain native-like environment for biomolecules and cells, and to control tip-surface forces and eliminate capillary interactions, thereby reducing sample damage. Conventional electrochemical SPMs typically operate in a very low frequency ranges (dc to ~ kHz) and require control of the probe and counter electrode potentials with ~mV resolution. In contrast, PFM operates at 1 kHz – 5 MHz frequency range and typically requires large (0.3 – 20 $V_{pp}$) modulation voltages to measure small (~3 – 500 pm/V) electromechanical coupling coefficients. Implementation of PFM in liquid potentially allows (I) operation in the media compatible with biological systems, (II) precise control of tip-surface interactions through the elimination of capillary forces and control over Van der Waals interactions, which largely eliminates jump-to-contact instability, (III) reduction of electrostatic interactions through matching of dielectric constants between solvent and substrate. However, factors such as (1) electrochemical reactions, (2) liquid effect on cantilever dynamics through added mass and fluid damping, and (3) stray currents, can significantly affect signal formation mechanism and limit potential for PFM imaging in liquid environment. This broad range of contributing factors precludes unambiguous theoretical determination of the likelihood for successful electromechanical imaging by PFM in liquid.

Recently, we have demonstrated PFM imaging of a model ferroelectric system (lead zirconate-titanate) in a conductive liquid environment [36]. It was demonstrated that the use of high modulation frequencies (1-2 MHz) largely precludes electrochemical processes even for large biases (10 $V_{pp}$) and minimizes liquid damping and added mass effects. Screening by mobile ions reduces the spatial extent of electrostatic fields, and allows imaging at high resolutions. Here, we present initial results on electromechanical imaging of a variety of biosystems in a liquid environment and analyze likely sources of contrast in liquid PFM images. Finally, we discuss the technical challenges, theoretical considerations, and expected outcomes for PFM in liquid environments.

## II. EXPERIMENT

### II.1. Principles of PFM

Piezoresponse force microscopy is based on the detection of the bias-induced piezoelectric surface deformation. The tip is brought into contact with the surface, and the piezoelectric response of the surface is detected as the first harmonic component, $A_{1\omega}$, of the tip deflection, $A = A_0 + A_{1\omega}\cos(\omega t + \varphi)$, induced by the application of the periodic bias, $V_{tip} = V_{dc} + V_{ac}\cos(\omega t)$, to the tip. The phase of the electromechanical response of the surface, $\varphi$, yields information on the sign of the piezoresponse.

PFM is implemented on commercial SPM systems (Veeco NS-IIIA MultiMode, NanoMan V, and Asylum Research MFP-3D) equipped with additional function generators and lock-in amplifiers (DS 345 and SRS 830, Stanford Research Instruments, and Model 7280, Signal Recovery). Measurements in ambient were performed using Pt and Au coated tips (NSC-12 C, Micromasch, $l \approx 130$ μm, resonant frequency ~ 150 kHz, spring constant $k$ ~ 4.5 N/m). Measurements in liquids were performed using Cr-Au-coated CSC-38 cantilever a with $l \approx 300$ μm, resonant frequency ~ 14 kHz, and spring constant $k$ ~ 0.05 N/m. Vertical

PFM (VPFM) measurements were performed at frequencies 50 kHz – 1 MHz, which minimizes the longitudinal contribution to measured vertical signal and allows imaging close to the resonances [37].

**II.2. Model Systems**

II.2.1. Lysozyme fibrils

Fibril formation and purification procedures were based on those reported by Vernaglia *et al*. [38]. A 6 mg/mL chicken egg-white lysozyme (lysozyme, #L6876 from Sigma Chemical Co.) stock solution was prepared in 20 mM potassium phosphate ($KH_2PO_4$), adjusted to pH 6.3 ± 0.1 with 1 M potassium hydroxide. Dilution with 4 M guanidine hydrochloride (GuHCl, #105696 from MP Biomedicals, Inc.) in 20 mM $KH_2PO_4$ yielded ~1.8 mg/mL lysozyme in ~2.8 M GuHCl and 20 mM $KH_2PO_4$. Fibrillization was induced by incubating the solution for 24 hours in a glass vial on a heated stir plate at 50°C with mild agitation (~100 rpm). The fibrils were centrifuged at 14,100 *g* for 10 minutes, the supernatant was discarded, and the pellet was washed with the same volume of HPLC-grade water (Fisher Chemical) using gentle shaking for 10 minutes. This step was repeated for a total of 15 washes. The fibril suspension was diluted with HPLC-grade water to 0.25 mg/mL, and samples for PFM imaging were prepared by incubating 25 μL of the resulting suspension on freshly cleaved mica (#71851-05 from Electron Microscopy Sciences) at room temperature for 2.5 minutes. Incubation was followed by 30 washes with 100 μL HPLC-grade water, and gentle drying with compressed nitrogen. Another 20 washes with 100 μL HPLC-grade water and finally gentle drying with compressed nitrogen completed the adsorption of lysozyme onto mica. Before imaging, 100 µL of HPLC-grade water was added onto the sample for solution PFM imaging.

II.2.1. Insulin Amyloid fibrils

Insulin amyloid fibrils were formed and purified following the procedures reported by Guo and Akhremitchev [39]. Bovine insulin (#I5500 from Sigma-Aldrich) was reconstituted to 5 mg/mL in 10mM hydrochloric acid (HCl, Fisher Chemical). This solution was incubated in a polypropylene microcentrifuge tube in a block heater at 80°C for 48 hours. The fibril suspension was diluted 10-fold with 10mM HCl to 0.5 mg/mL protein and was centrifuged at 3,000 $g$ for one minute to remove smaller aggregates from the longer fibrils. The precipitates were collected and washed by gentle shaking for five minutes with the same volume of 10mM HCl as had been removed. This was repeated two times for a total of three washes. The fibril suspension was diluted 1:1 with 10 mM HCl to 0.25 mg/mL, and samples for PFM imaging were prepared by incubating 25 µL of the resulting suspension on freshly cleaved mica (#71851-05 from Electron Microscopy Sciences) at room temperature for one minute. Incubation was followed by 5 washes with 35 µL of 10 mM HCl and gentle drying with compressed nitrogen. The residual HCl was then removed by five additional washes with 35 µL of HPLC-grade water and finally gentle drying with compressed nitrogen. Before imaging, 100 µL of HPLC-grade water was added onto the sample for solution PFM imaging.

II.2.3. Breast adenocarcinoma cells

Breast adenocarcinoma cells (MCF7 cell line) were cultured in Dulbecco's Modified Eagle's Medium (DMEM) with glucose and L-glutamine (#D5523 from Sigma-Aldrich) supplemented with 10% FBS (#ES1055 from Biomeda), Penicillin-Streptomycin (#P0781 from Sigma-Aldrich), and non-essential amino acids (#M7145 from Sigma-Aldrich). The cells were plated onto pretreated Corning 60mm polystyrene Petri dishes (#430166 from Fisher Scientific) and were maintained in a CO2 incubator at 37 ºC until imaging.

II.2.4. Bacteriorhodopsin

Bacteriorhodopsin from *Halobacterium salinarium* (#B0184 from Sigma-Aldrich) was diluted to a final imaging concentration of 50μg/ml in 10mM Tris, 150mM KCl buffer, pH 8. To prepare the sample, 10μl of the bacteriorhodopsin solution was added to 10μl of buffer onto a freshly cleaved mica disc. The solution was incubated for 10 minutes then rinsed with 100μl of buffer solution. Additional buffer was added for a final volume of 200μl. Patches of bacteriorhodopsin were imaged with a metal coated tip in tapping mode with low drive and setpoint amplitudes.

### III. Results

### III.1. Liquid PFM imaging of model ferroelectric systems

To illustrate the feasibility of electromechanical imaging in liquid environment, the PFM was performed in air and in DI water on a periodically poled lithium niobate (PPLN, Crystal Technology) as a model system. The electromechanical response of PPLN in the point contact geometry of a PFM experiment is well-known both from calculations [40] and direct measurements [41] and is ~12 pm/V. PPLN is thus a convenient model system for calibration of the system, i.e. determination of the conversion factor between the lock-in amplifier output and local electromechanical response. Furthermore, the noise level in the PFM measurements can be determined from the dispersion of the signal on a single domain. If the vertical sensitivity of the system is known independently, the electrical contact quality between the tip and the surface can be estimated. Note that both types of calibration are generally applicable only for frequencies well below the first resonance of the cantilever. Otherwise, cantilever dynamics which is strongly dependent on the contact stiffness of tip-surface junction, will provide an incontrollable contribution to the signal [42].

The PFM amplitude and phase images taken from the PPLN in air and distilled water are shown in Fig 1. Both images demonstrate clear domain contrast unrelated to topographic features in amplitude images. The phase contrast in both cases illustrates phase change by 180° between antiparallel domains. The noise level and effective domain wall width in liquid image are higher, indicative of more significant spreading of the electric field anticipated in the high dielectric constant ($\varepsilon = 80$ in water vs. $\varepsilon = 1$ in air). However, clear qualitative agreement between PFM images in both environments serves as a strong indicator of the electromechanical nature of the contrast.

### III.2. Liquid PFM imaging of amyloid fibrils

After verification on the ferroelectric system with known electromechanical response, PFM was applied to a variety of biological systems. Figure 2 shows topographic (a,d,g) and PFM images of a lysozyme fibril as a function of imaging bias, $V_{ac}$. PFM phase images are shown in (b,e,h) and PFM amplitude images are shown in (c,f,i) for 10 V, 5 V, and 2 V, respectively. The piezoresponse amplitude linearly depends on the applied bias, and the phase response is stronger for higher bias. The fibril does not appear to degrade with repeated scanning or application of ac bias. Some internal structure is discernable in PFM amplitude images, while the phase response appears to be uniform within a fibril. Similar behavior is observed for the insulin fibril in Fig. 3, shown for two driving voltages. The strong dependence of the phase contrast on driving amplitude and small phase shifts (<<180°) between dissimilar regions suggests significant contribution of the electrostatic forces (modified by the presence of liquid) to the PFM signal. Based on the visual inspection of the topographic, PFM phase and PFM amplitude data it is obvious that (1) the apparent spatial resolution in the PFM amplitude image is higher compared to topography and (2) phase images show clear contrast on the parts of the fibril in which topographic height is on the limit of detectability. Below, we explain that these observations by assuming that the main electromechanical signal is related to the variability of the local elastic properties between the substrate and the fibril. This elastic contrast is convoluted with electromechanical and electrostatic responses, precluding unambiguous determination of the latter. However, increased resolution and detection limit clearly illustrate the usefulness of this detection method.

### III.3. Imaging of cellular systems

The extension of liquid PFM imaging to viable cells is illustrated in Fig. 4, showing topography, PFM amplitude and PFM phase of living breast adenocarcinoma cells. Only weak contrast between different cells or at cell boundaries is observed in the PFM amplitude image. However, the cells with higher height profiles appear to have a slightly different PFM phase response than the surrounding cells. In addition, cell boundaries are resolved in the PFM phase image. Similarly to the amyloid and insulin fibrils, these observations suggest the possibility of strong elastic contribution to the PFM signal. The reason for relatively poor quality of cell PFM images is the presence of electrolytes in the cell culture media because high conductivity of the solution results in the uniform biasing of the liquid. To avoid this problem, the imaging should be performed using insulated or shielded probes, in which the central part of the tip is insulated form the solution except for the region near the probe apex.

### III.4. Electromechanical imaging of bacteriorhodopsin and PFM resolution

The general analysis of PFM image formation suggests that the electromechanical contribution to the signal is favored by high effective stiffness of the system. This is due to the fact that electromechanical contribution is independent of spring constant, and electrostatic contribution scales reciprocally with effective spring constant.[42] Furthermore, optimal imaging conditions with minimal topographic cross-talk correspond to flat surfaces. Hence, we have chosen bacteriorhodopsin (BR) purple membrane as a relatively stiff and topographically flat model system for biological system. Furthermore, strongly asymmetric structure of the BR with significant intrinsic dipole moment suggests the potential electromechanical response. Shown in Fig. 5 are the topographic and PFM amplitude images of the BR. In this case, imaging was performed using ac mode (mechanical tip oscillation at first resonance) to obtain topographic data. Electrical modulation was performed at the second resonance for amplitude [43]. The top surface of the membrane exhibits islands with double height, corresponding to membrane folding. The PFM amplitude image shows the signal of equal strength within the membrane surface and the islands, with characteristic depression at the boundary. At the same time, phase image shows signal inversion between the island and the membrane. This behavior is qualitatively similar to that expected for ferroelectric or piezoelectric domains. Based on these observations, and given minimal potential for topographic or elastic cross-talk on the uniform surface, we attribute the observed contrast to the difference in electromechanical response (e.g. opposite sign of piezoelectric coefficients) between the extracellular and cytoplasmic regions of the BR.

The experiments with BR also establish the potential for molecular-level spatial resolution. The operation of PFM requires reliable control of both dc and ac tip bias in the 1 kHz – 1 MHz range. Operation in liquid environment brings additional limitations on the conductance of the tip coating, which should be greater than that of the liquid to stray current paths through the solution. This necessitates the use of metallic or metal coated tips, which typically have significantly higher radii of curvature compared to Si or $Si_3N_4$ probes. Figure 6 illustrates that the use of intermittent contact mode with a metal coated tip in liquid allows molecular resolution to be achieved.

## IV. Towards Electromechanical Imaging in Liquids

The results summarized in Section III illustrate that electromechanical imaging in liquid environment produces measurable contrast on a variety of biological systems. The presence of the contrast *per se* evidences measurable electromechanical and electrostatic interactions between the probe and the surface in liquid. In many (albeit not in all) cases, the contrast is visually distinct from surface topography or its first- and second derivatives, suggesting that it is unrelated or only partially related to topographic cross-talk. Furthermore, contrast scales linearly with probing voltage, consistent with its (linear) electromechanical [e.g. piezoelectric or flexoelectric] origins. Based on these observations, we analyze the possible origins of observed signal and summarize the requirements for quantitative studies of electromechanical activity in liquids.

Local quantitative mapping of electromechanical activity in liquid environment requires to:

1. Concentrate the probing ac field
2. Concentrate and/or control local electrochemical potential through the dc field
3. Detect local surface displacement induced by ac potential (electromechanical response)
4. Separate electromechanical response from electrostatic capacitive interactions
5. Avoid or separate topographic cross-talk that can contribute to measured signal

The theory of PFM, including electrostatic and strain field structure induced by the probe, resolution theory, local and non-local contributions to signal, surface-tip signal transfer and its frequency dispersion, are well studied in the context of ambient PFM [44]. Below, we briefly discuss principles of PFM imaging in ambient environment, and analyze some of the factors that can affect imaging in liquids.

**IV.2. Force interactions in liquid and air**

The key element of any scanning probe microscopy (SPM) method is the mechanism of tip-surface interactions that determine the strength of the force signal detected by the cantilever, locality of the imaging, and primary response mechanism. In the context of PFM, primary interactions include electrostatic interactions and electromechanical coupling in the material and electric double layers. Experimentally measured piezoresponse amplitude is $A = A_{el} + A_{piezo} + A_{nl}$, where $A_{el}$ is electrostatic contribution, $A_{piezo}$ is electroelastic contribution and $A_{nl}$ is non-local contribution due to capacitive cantilever-surface interactions [45,46]. Quantitative PFM imaging requires $A_{piezo}$ to be maximized to achieve predominantly electroelastic contrast. The origins of electromechanical contrast and its relationship to local materials properties for piezoelectric materials have been extensively studied [47,48]. Piezoelectric coupling is specific to materials, and thus is not expected to be affected by the environment. At the same time, electrostatic force interactions are strongly mediated by medium.

In ambient environment or in vacuum, the tip interacts with the surface through long-range electrostatic forces, $F_{el}^{amb} = C_z'(z)(V_t - V_s)^2$, where $V_t$ is tip potential, $V_s$ is surface potential, and $C_z'(z)$ is tip-surface capacitance gradient. In the low-frequency limit, the piezoresponse (PR) signal, i.e. the first harmonic component of tip oscillations induced by periodic bias applied to the tip, is:

$$PR_{amb} = \alpha_a(h)\tilde{d}_{33}\frac{k_1}{k_1+k} + \left(\frac{C_{sphere}' + C_{cone}'}{k_1+k} + \frac{C_{cant}'}{24k}\right)(V_{dc} - V_s), \quad (1)$$

where $\tilde{d}_{33}$ is the effective electromechanical response of material, $\alpha_a(h)$ is the ratio of ac tip potential to the ac surface potential (i.e., the potential drop in the tip-surface gap), $k_1$ is the spring constant of the tip-surface junction, $k$ is the spring constant of the cantilever, $C_{sphere}'$, $C_{cone}'$, and $C_{cant}'$ are capacitance gradients due to the spherical and conical part of the tip and cantilever. $V_{dc}$ is the dc potential offset of the tip bias, and $h$ is the tip-surface separation. In ambient $\alpha_a(h) = 1$ for $h < 0$ (contact), i.e., the response is independent of penetration depth, and $\alpha_a(h) \ll 1$ for $h > 0$ (non-contact), giving rise to well-known limits of PFM and Kelvin probe force microscopy.

In liquid, the PFM contrast is strongly mediated by the presence of mobile ions that screen electrostatic tip-surface interactions. For the sphere-plane system [49],

$$F_{el}^{l}(z) = \frac{\varepsilon\varepsilon_0 R}{\lambda_D} \frac{2V_t V_s \exp(h/\lambda_D) - (V_t^2 + V_s^2)}{\exp(2h/\lambda_D) - 1}. \tag{2}$$

Electrostatic interaction in liquid is short range and decays exponentially for $h > \lambda_D$. Thus, the PFM signal in liquid (for $h > 0$) is

$$PR_l = \alpha(h) d_{33} \frac{k_1^l}{k_1^l + k} + \frac{\varepsilon\varepsilon_0 R}{\lambda_D} \frac{2V_s \exp(h/\lambda_D)}{\exp(2h/\lambda_D) - 1} \frac{1}{k_1^l + k}, \tag{3}$$

where the first term is piezoelectric response, $k_1^l$ is tip-surface spring constant in liquid, and the second term is the response due to the electrostatic coupling in the double layer. The contribution from the conical part of the tip and cantilever are absent for $R \gg \lambda_D$. The screening coefficient $\alpha(h) = 1$ for $h \ll \lambda_D$, i.e. when the tip touches the surface, and is $\alpha(h) = 0$ if the electric double layers around the tip and the surface do not overlap. Thus, the electromechanical response in solution gradually decays on the distances on the order of the Debye length of the solution, and the electrostatic contribution is significantly minimized compared to ambient or vacuum imaging.

### IV.3. AC and DC Field Concentration

The second key component of the PFM imaging is the concentration of the ac and dc field in the tip-surface junction and detection of local response. Notably, simple observation of high-resolution contrast is not enough to demonstrate the high localization of an ac electric field. Rather, locality of the strain transfer through the mechanical tip-surface contact is a sufficient explanation. In the absence of additional information, the relative size of electrical (ac field region) and mechanical contact can not be established and the spatial resolution provides only the estimate of the lower of the two.

The distribution of the dc field however can be probed directly by the polarization switching experiments in the liquid environment. A transition of ferroelectric switching behavior from localized to uniform switching depending on the choice of the solvent established that imaging is possible at conductivities far larger than allowed for localized switching [50]. Furthermore, these results illustrated the degree to which the spatial extent of a dc field can be controlled in solution. Application of dc pulses is possible only in less conductive non-aqueous solvents using conventional metal coated cantilevers.

### IV.4 Contact mechanics

The third element determining PFM response is the signal transfer from material to the tip, determined by the cantilever spring constant, $k$, and tip-surface spring constant, $k_1$. The importance of variations of effective spring constant on signal directly follows from Eq. (1), in which variations in local stiffness couple to electromechanical signal through the capacitive forces. This coupling will be the origin of both topographic and elastic cross-talks through variations in tip-surface contact stiffness.

These effects can be analyzed using a simple Hertzian model for the tip-surface contact. The relationship between the indentation depth, $h$, tip radius of curvature, $R_0$, and load, $P$, is [51]

$$h = \left(\frac{3P}{4E^*}\right)^{\frac{2}{3}} R_0^{-\frac{1}{3}} \qquad (4)$$

where $E^*$ is the effective Young's modulus of the tip-surface system. The contact radius, $a$, is related to the indentation depth as $a = \sqrt{hR_0}$. The contact stiffness is $k_1 = (\partial h/\partial P)^{-1}$, and from Eq. (42), $k_1 = 2aE^*$, or

$$k_1 = 2E^* \sqrt{hR_0} = \left(6PE^{*2}R_0\right)^{\frac{1}{3}} \qquad (5)$$

Shown in Fig. 7 (a) is the distance dependence of the electrostatic and electromechanical contributions to the PFM signal calculated for $R = 50$ nm, $V_{dc} = 0.1$ V, $E^* = 100$ GPa, $d_3 = 50$ pm/V, and $k = 1$ N/m and 40 N/m. The distance dependence of the capacitive tip-surface forces and tip-induced surface potential was calculated neglecting the changes in the sphere area due to contact. The electrostatic contribution decreases rapidly with penetration depth due to changes in the tip-surface stiffness constant. In comparison, shown in Fig. 7 (b) is the fraction of the electromechanical contribution depending on penetration depth. Note that in the Hertzian model, the electromechanical signal dominates for a penetration depth of ~1 A, corresponding to a contact radii on the order of ~2 nm for $R = 50$ nm, imposing a limit on the spatial resolution of the technique.

The analysis becomes more complicated if adhesive effects are taken into account. In this case, the contact mechanics are described by the Johnson-Kendall-Roberts (JKR) model, applicable for ambient conditions. In the JKR model, the contact radius is

$$a^3 = \frac{R_0}{E^*}\left(P + 3\sigma\pi R_0 + \sqrt{6\sigma\pi R_0 P + (3\sigma\pi R_0)^2}\right) \qquad (6)$$

where $\sigma$ is the work of adhesion, $P$ is the load and indentation depth is

$$h = \frac{a^2}{R_0}\left[1 - \frac{2}{3}\left(\frac{r_0}{a}\right)^{3/2}\right] \qquad (7)$$

where $r_0^3 = 6\sigma\pi R^2/E^*$ is contact radius at zero force. Shown in Fig. 8 (a) are force vs. indentation depth curves calculated for $\sigma = 0$ (Hertzian), $10^{-3}$, $10^{-2}$, $10^{-1}$, and 1 J/m². Shown in Fig. 8 (b,c) are corresponding contact stiffnesses. Note that adhesive contact results in rapid change of contact stiffness from 0 to the value corresponding to contact, resulting in a well-defined boundary between free and bound cantilevers. Finally, shown in Fig. 8 (d) is the force dependence of the contact area. Even for a small work of adhesion, the contact radii at zero force are relatively large, on the order of nanometers. Corresponding contact stiffnesses are of the order of 100–1000 N/m, well above the typical spring constant of the cantilever.

Thus, for most cantilevers (0.01 – 50 N/m), jump-to-contact instability in air distinguishes the regions with $k_1 \ll k$ [free air, $k_1 = 0$], and $k_1 \gg k$ [in contact $k_1 \sim 1000$ N/m]. These regions correspond to the purely electrostatic interaction in the Kelvin Probe Force Microscopy (KPFM), and predominantly electromechanical contrast in PFM. Conversely, in liquid the lack of capillary forces and reduced magnitude of Van der Waals (VdW) forces suggests that $k_1$ can vary continuously when approaching the surface. Hence, there is no well-defined boundary between predominantly electromechanical and purely electrostatic contrasts, and hence PFM and KPFM regimes are not well defined. Notably, similar behavior is anticipated for the non-contact AFM and KPFM in ultrahigh vacuum.

In both Hertzian and JKR models, the transition to the predominantly electromechanical contrast occurs for contact areas larger than a certain critical value. From Eq. (1), when non-local electrostatic components are negligible, this condition can be generalized for materials with arbitrary properties as

$$a > a^* = \frac{C'_{sphere}(V_{dc} - V_s)}{2\alpha(h)d_3 E^*} \qquad (8)$$

where $a^*$ is the critical contact radius corresponding to equality of the electrostatic and electromechanical contributions to the signal. For classical ferroelectric materials (100 GPa, 50 pm/V), this conditions becomes $a > a^* = 5(V_{dc} - V_s)$ A/V. For soft materials, (10 GPa, 5 pm/V), $a > a^* = 50(V_{dc} - V_s)$ nm/V.

From this simple estimate, the resolution of PFM on hard materials can potentially achieve sub-nanometer resolution provided that the electrostatic contribution to the signal is minimized. For soft systems with small response coefficients (~1-100 pm/V), the signal is likely to represent the convolution of electrostatic and electromechanical contributions. At the same time, for system with large electromechanical responses (i.e. many cells have ~$10^4$ higher electromechanical coupling than inorganic piezo- and ferroelectrics) electromechanical response is observable.

### IV.5. Topography effect (cross-talk)

During imaging of a variety of materials, or objects with high aspect ratios, care must be taken to minimize topographic cross-talk with the electromechanical response through electrostatic interactions and variations in contact stiffness. These effects are expected to be particularly important when imaging at frequencies close to contact resonances of the tip. In this case, even minute changes in local properties and topography can flip the system through the resonance, with associated 180° change in phase response and spurious changes in the amplitude. From the structure of Eq. (1), the following contributions can be differentiated:

- Variations of long-range electrostatic interactions due to capacitive terms dependence on topography, $C'_{sphere} + C'_{cone}$, and local work functions, $V_s$
- Variations of effective electromechanical response with topography, $\tilde{d}_{33}$
- Variation of electrical contact quality, $\alpha_a(h)$
- Variations of local contact stiffness, $k_1$

Topographic effects on scanning Kelvin probe microscopy, a non-contact (weakly dependent on contact area), lift-mode technique, have revealed that forces are larger within topographic depressions, but that these relatively weak long-range effects often show up as offsets in the signal [52]. Therefore, capacitive terms are unlikely to produce sharp cross-talk contrast on the 1-100 nm length scale.

The detailed studies of the voltage-dependent contact mechanics of piezoelectric materials, as well as simple dimensionality arguments, suggest that electromechanical response is only weakly dependent on surface topography. The electrical contact quality is expected to be only weakly dependent on contact area for a good contact (fringing field in capacitor), but is highly sensitive to surface contamination.

Finally, the most significant contribution in PFM cross-talk is the effect of topography on contact stiffness, resulting in strong cross-talk with topography in the resonance-enhanced PFM. This will result in the fine structure on the image, since resolution in this case is determined by the tip-surface contact area and large-scale variations in topography do not contribute to signal. The magnitude of this effect is studied in detail elsewhere, and it is shown that due to the specificity of PFM driving mechanism, the stable imaging in the vicinity of the resonance can be achieved only using fast sweeps, sampling a segment of the amplitude-frequency curve in the vicinity of the resonance [53]. This limitation is especially pronounced for weakly piezoelectric materials such as calcified and connective tissues with small (1-5 pm/V) coupling coefficients.

## V. Summary

We have studied in liquid the electromechanical properties of a variety of biological systems, including amyloid insulin fibrils, carcinoma cells, and purple membranes, as well as a ferroelectric surface for reference. Model ferroelectric system illustrates high-veracity electromechanical contrast clearly associated with domains that is observable even for relatively small (~12 pm/V) electromechanical response. For all biological systems, strong amplitude and phase contrast are observed in the PFM image. These data typically illustrate the level of detail exceeding the topographic resolution, and capacity for detection of signal even from small objects invisible on topographic image.

The PFM signal can originate from a variety of sources, including local electromechanical response, variations in contact stiffness due to topography and elasticity. In many systems, these sources of contrast cannot be distinguished unambiguously. However, studies on the purple membranes clearly demonstrate the difference in electromechanical response between the cytoplasmic and extracellular sides. This behavior is anticipated for the membrane with built-in dipole moment. Furthermore, these results illustrate that imaging with molecular resolution is feasible with metal-coated probes.

The significant bottle neck for future electromechanical studies in liquid is the need to concentrate both ac and dc electric field produced by the probe. This is required both to improve the resolution and to minimize the stray currents in liquid and electrochemical reactions. This necessitates the development of shielded probes AFM probes, as suggested by a number of groups [54-56].

Research sponsored in part (BJR and SVK) by the Division of Materials Sciences and Engineering, ORNL LDRD funding (BJR, SJ, and SVK) and ORNL SEED funding (SJ and SVK), Office of Basic Energy Sciences, U.S. Department of Energy, under contract DE-AC05-00OR22725 with Oak Ridge National Laboratory, managed and operated by UT-Battelle, LLC. Research was also supported through CNMS user proposals #2005-075 and #2006-049, and NSF #CMS-0619739 (GLT and AAV). The authors (BJR, SVK) are also grateful for financial support from Asylum Research and for the use of their imaging facilities.

**Figure Captions**

**Fig. 1** PFM (a,c) amplitude and (b,d) phase images of a periodically poled lithium niobate surface in air and distilled water, respectively. Data scales are 20 nm, 10 V, and 20 V for (a,b,c), and 50 nm, 20 V, and 10 V for (d,e,f), respectively.

**Fig. 2** PFM images of a lysozyme fibril adsorbed onto mica. The scan size is 2.97 µm x 2.97 µm, with a scan rate of 1 Hz and 512 samples per line. AC biases, 10 $V_{ac}$, 5 $V_{ac}$ and 2 $V_{ac}$, as shown in the top, middle and bottom rows, respectively, were applied with a frequency of 383 kHz, a 0° phase angle, and a lock-in bandwidth of 1 kHz. The first column shows topographic images, the second column - piezoresponse phase images, and the third column - piezoresponse amplitude images. Z scales are 50 nm, 360° and 2.5 mV for the topographic, piezoresponse phase and piezoresponse amplitude images, respectively. The height of the feature ranged from about 10 nm to a maximum of about 25 nm consistent with the diameters of lysozyme fibrils reported by Vernaglia *et al*. [38]. Note that the amplitude feature disappears with decrease in applied ac bias from 10 $V_{ac}$ (c) to 2 $V_{ac}$ (i) while the topographic and phase images reveal that the fibril bundle remains present.

**Fig. 3** PFM images of an insulin fibril adsorbed onto mica. The scan size is 603 nm x 603 nm, with a scan rate of 1 Hz and 512 samples per line. AC biases, 10 $V_{ac}$ and 2 $V_{ac}$, as shown in the top and bottom rows, respectively, were applied with a frequency of 437.8 kHz, a 0° phase angle, and a lock-in bandwidth of 1 kHz. The first column shows topographic images, the second column - piezoresponse phase images, and the third column - piezoresponse

amplitude images. Z scales are 25 nm, 180° and 2.0 mV for the topographic, piezoresponse phase and piezoresponse amplitude images, respectively. The height of the feature ranged from 5 to 10 nm, which is consistent with the range of insulin fibril heights reported by Guo and Akhremitchev [39]. Note that the amplitude feature decreases and fades with decrease in applied ac bias from 10 $V_{ac}$ (e) to 2 $V_{ac}$ (f) while the topographic and phase images reveal that the fibril bundle remains present.

**Fig. 4** Live breast adenocarcinoma cells (MCF7 cell line) imaged in DMEM culture medium. (a) Topography, (b) PFM amplitude, and (c) PFM phase images. Images were acquired in contact mode using the MFP-3D.

**Fig. 5.** Dual-ac mode images of bacteriorhodopsin imaged in buffer solution with a metal coated tip. Shown are (a) surface topography and (b) amplitude of the electromechanical response signal.

**Fig. 6** Topography images of bacteriorhodopsin acquired in buffer solution with a metal coated tip using the NanoScope MultiMode at different magnitfications. Z scales are 12 nm, 2 nm and 4 Å, respectively.

**Fig. 7** (a) Indentation depth dependence for electromechnical (blue) and electrostatic (red) contributions for a cantilever with k = 1 N/m (solid) and k = 40 N/m (dash). (b) Indentation depth dependence of the fraction of the electromechanical contribution.

**Fig. 8** (a) Indentation force-distance dependence in the JKR model. (b,c) force dependence of indentation stiffness. (d) Force dependence of contact radius. Shown are curves for $\sigma = 0$ (solid), $10^{-3}$ (dot), $10^{-2}$ (dash), $10^{-1}$ (dash-dot), and 1 J/m$^2$ (dash-dot-dot).

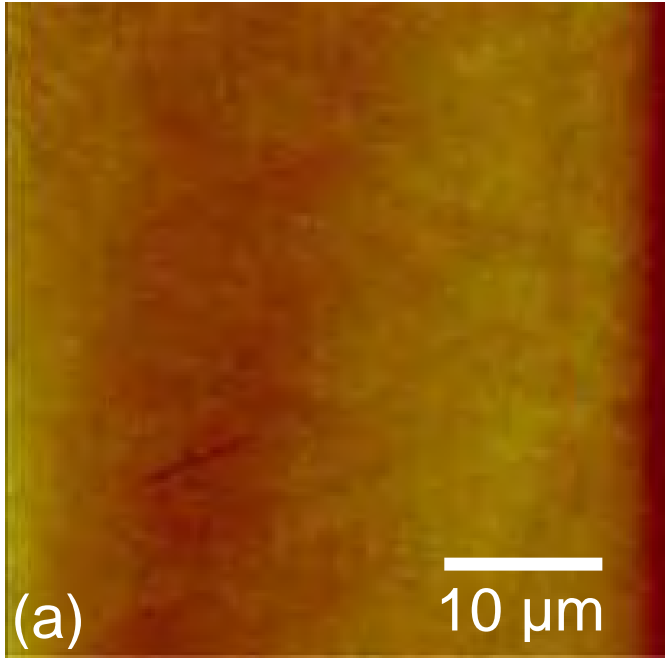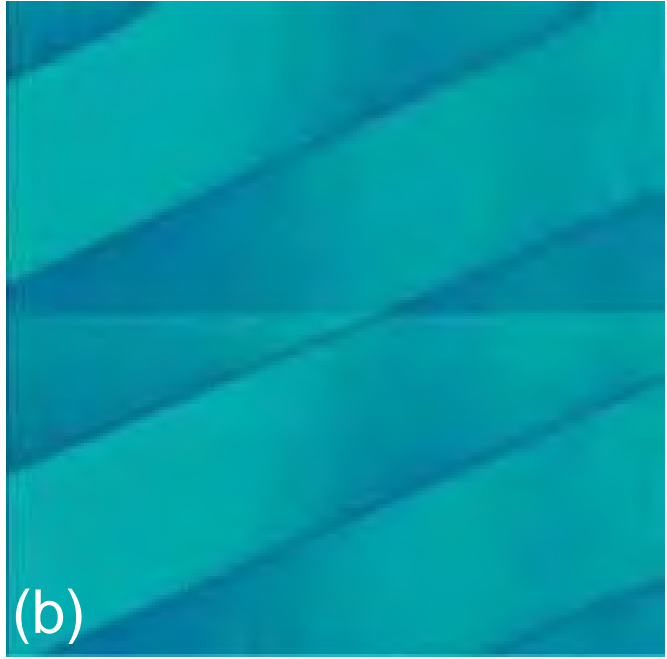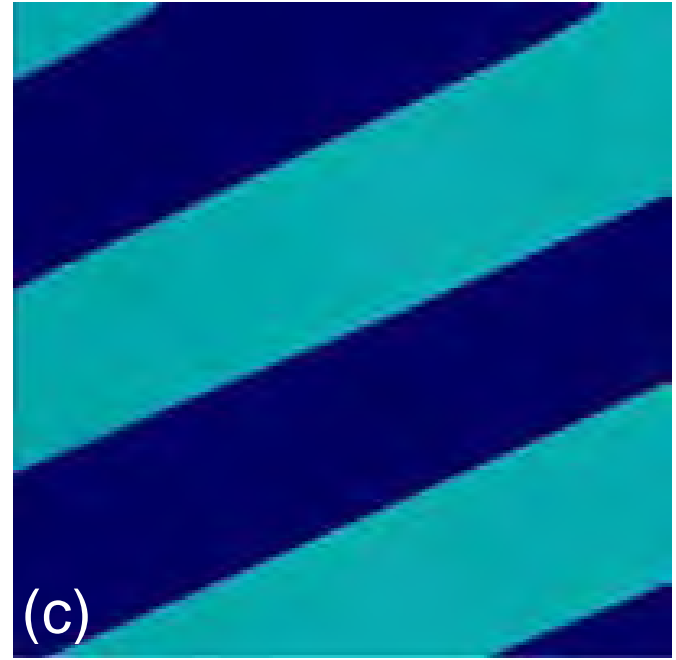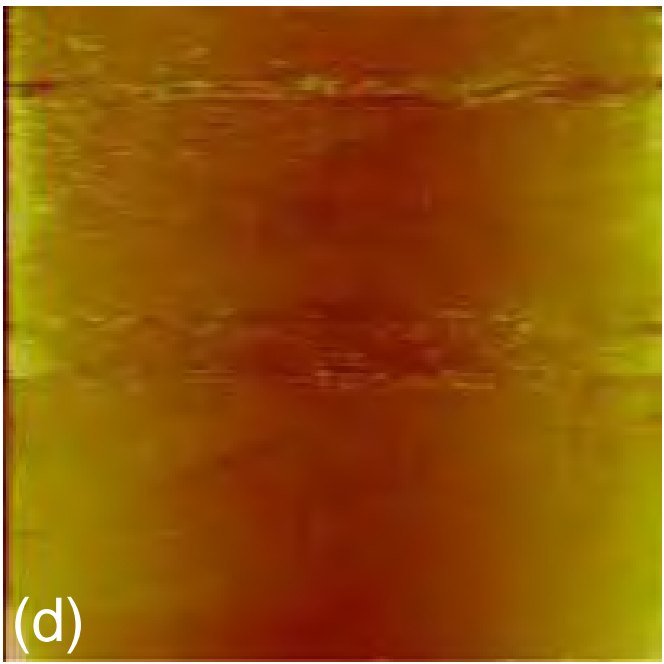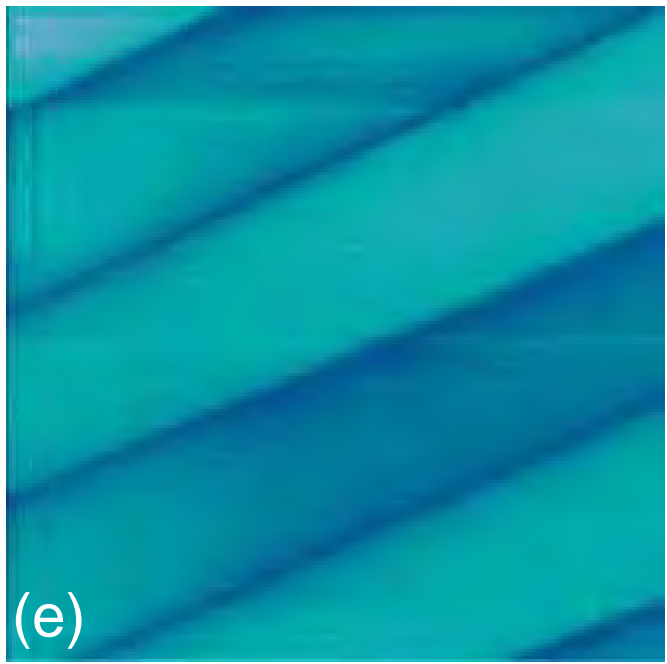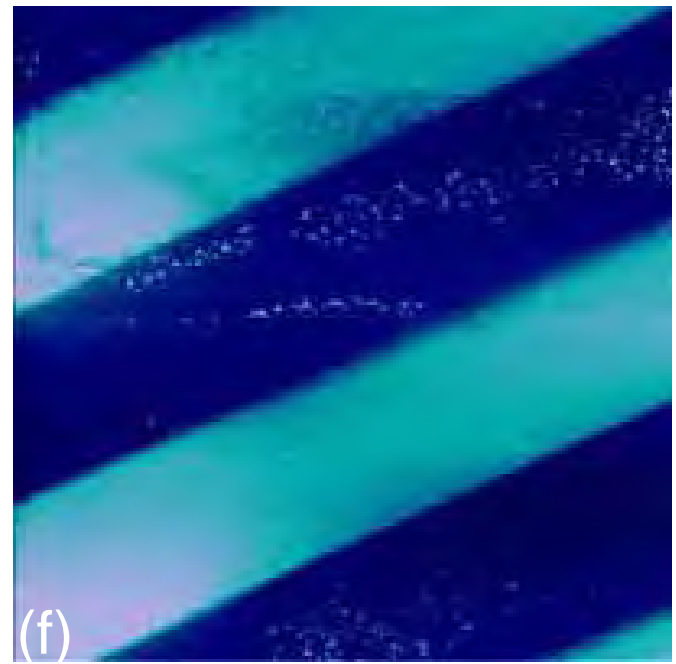

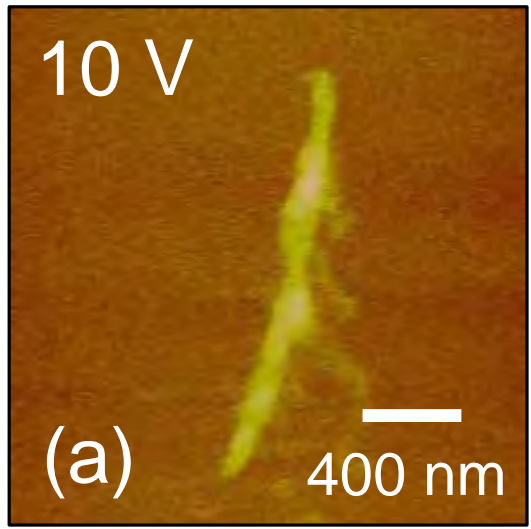 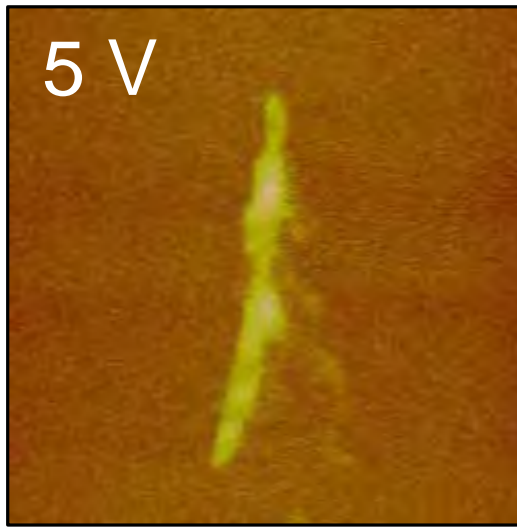 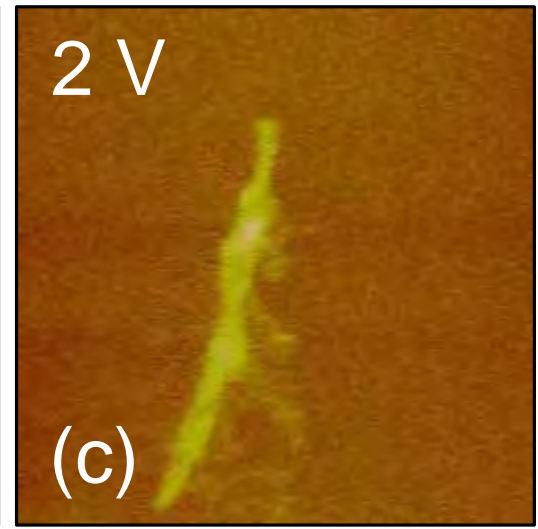
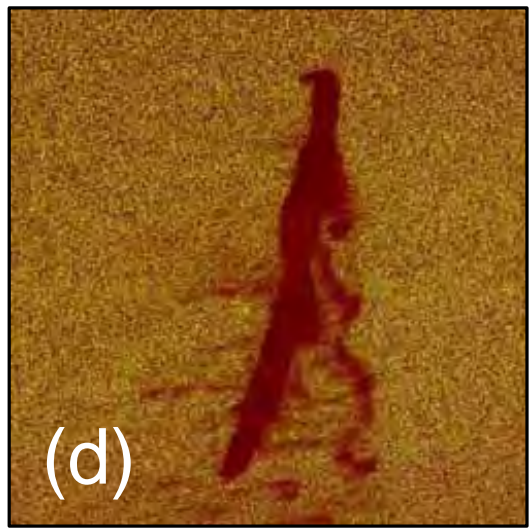 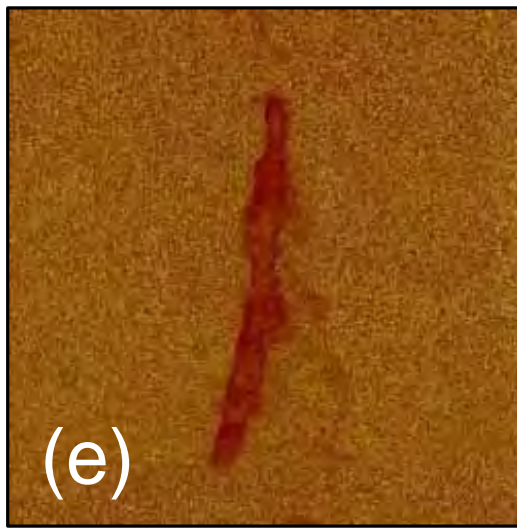 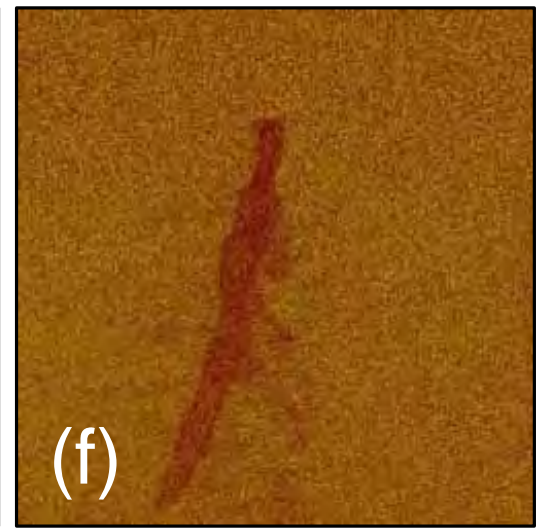
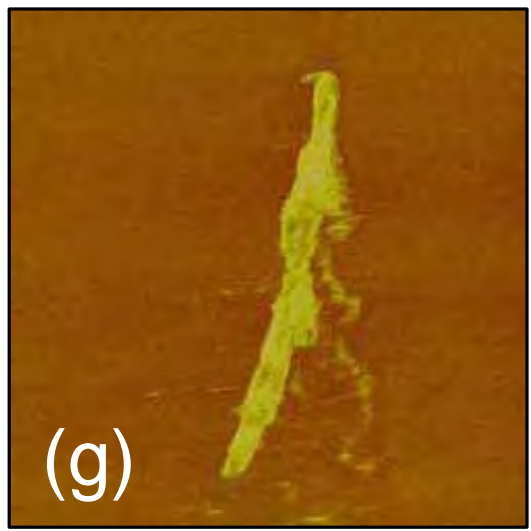 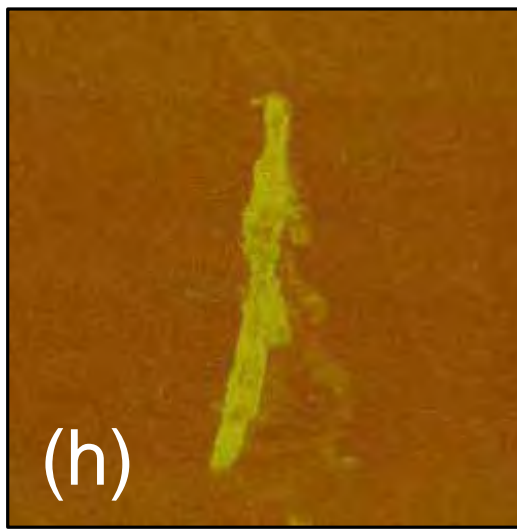 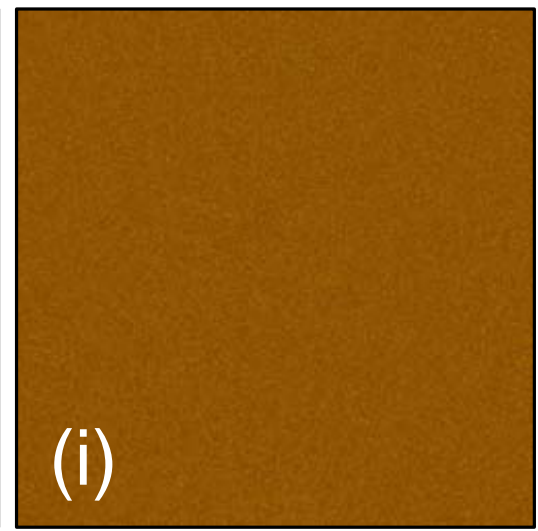

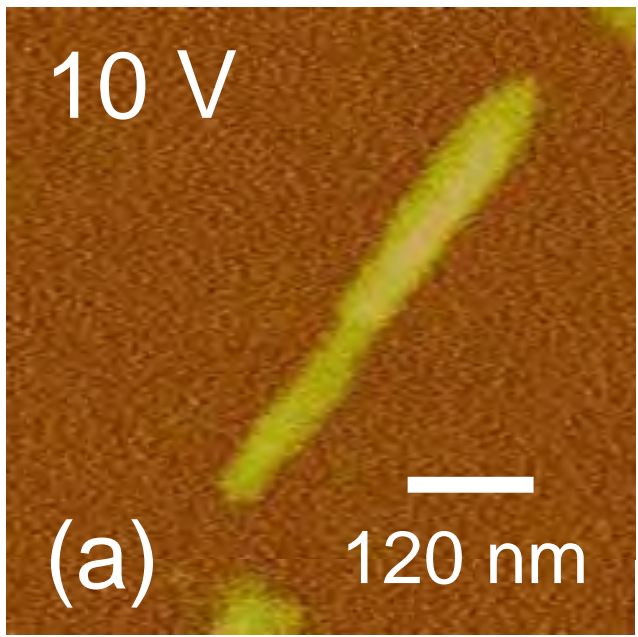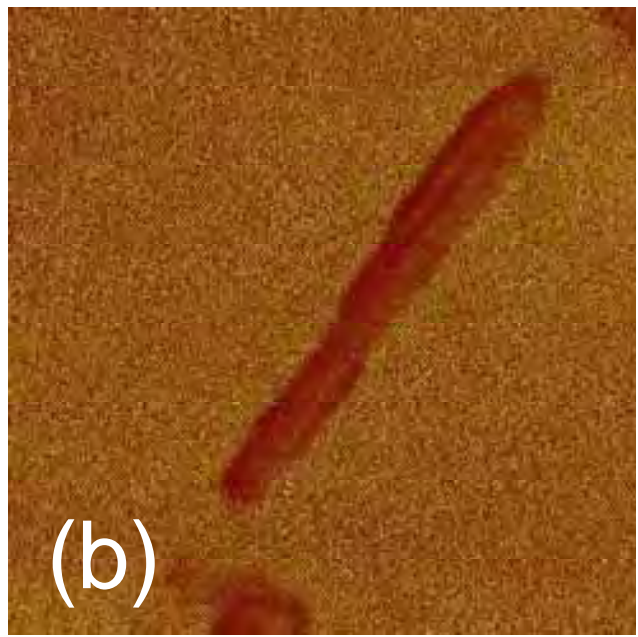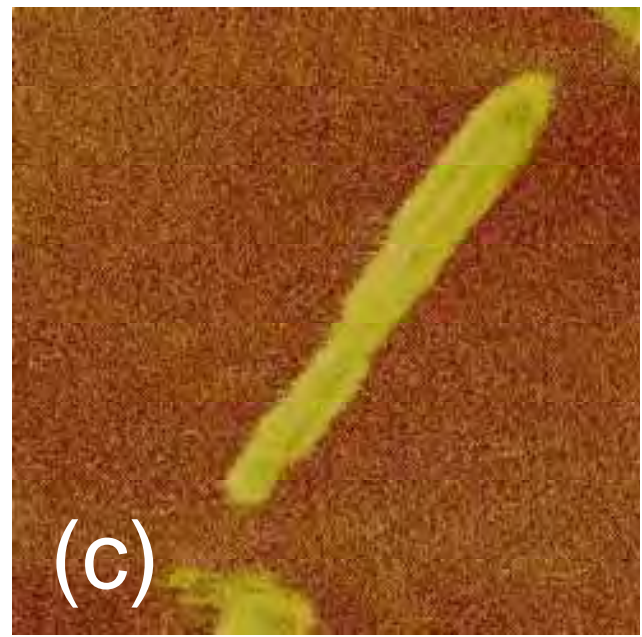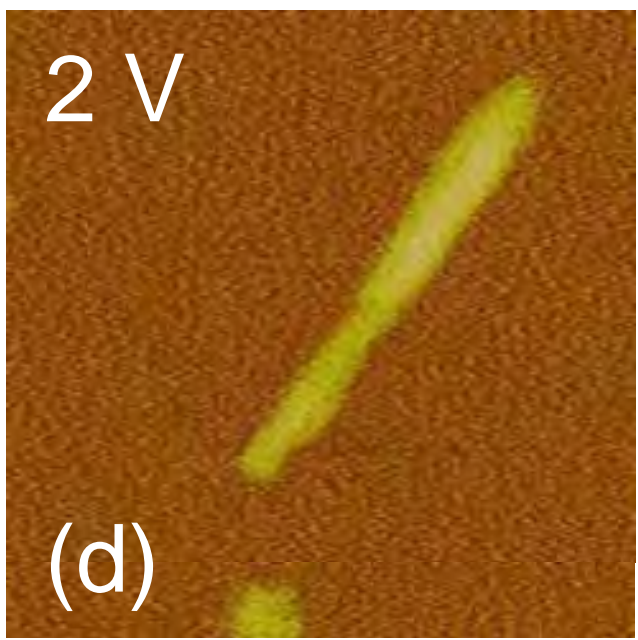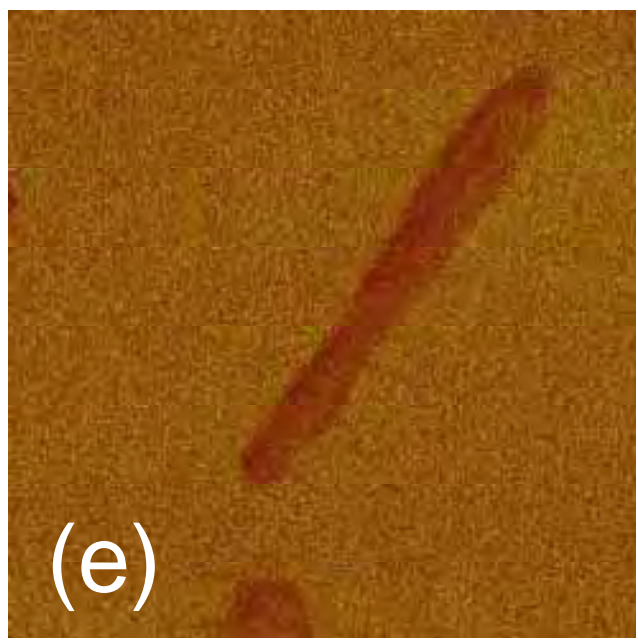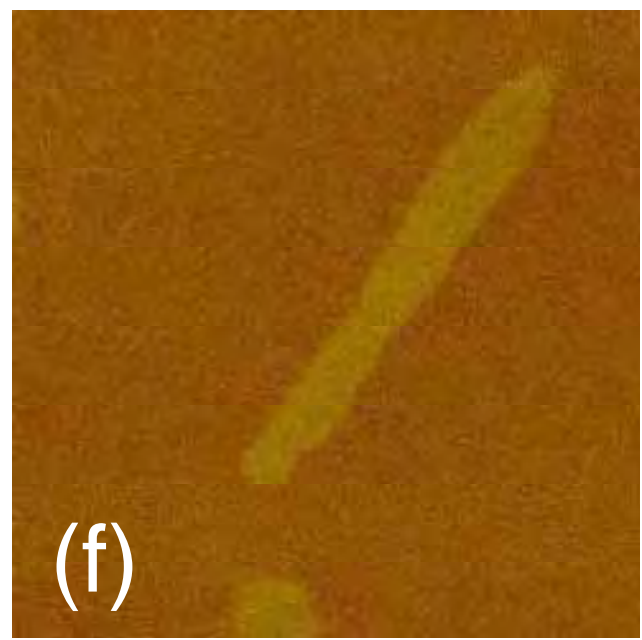

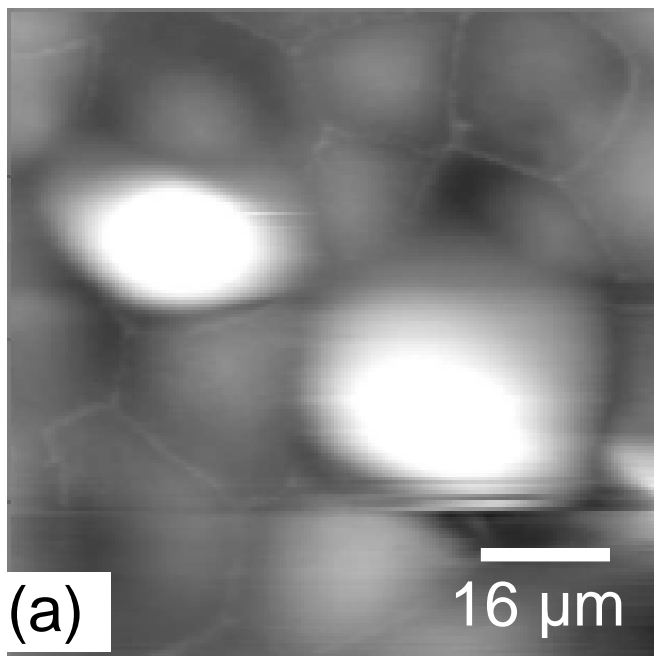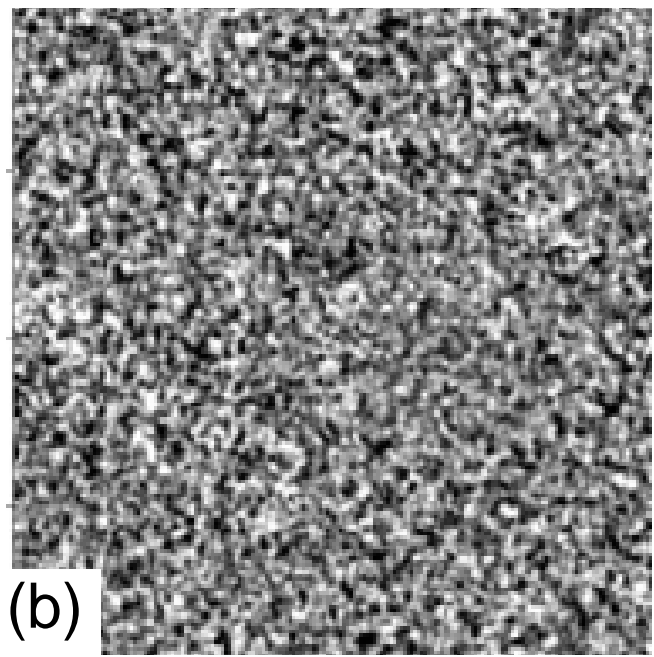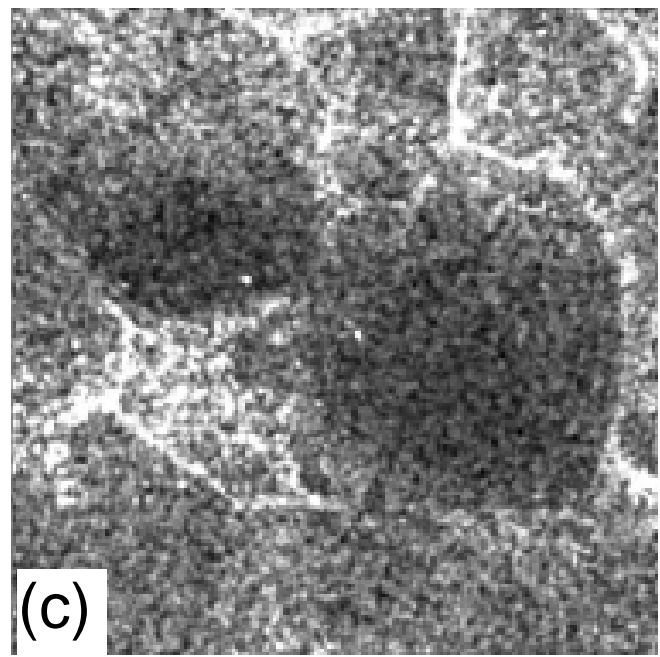

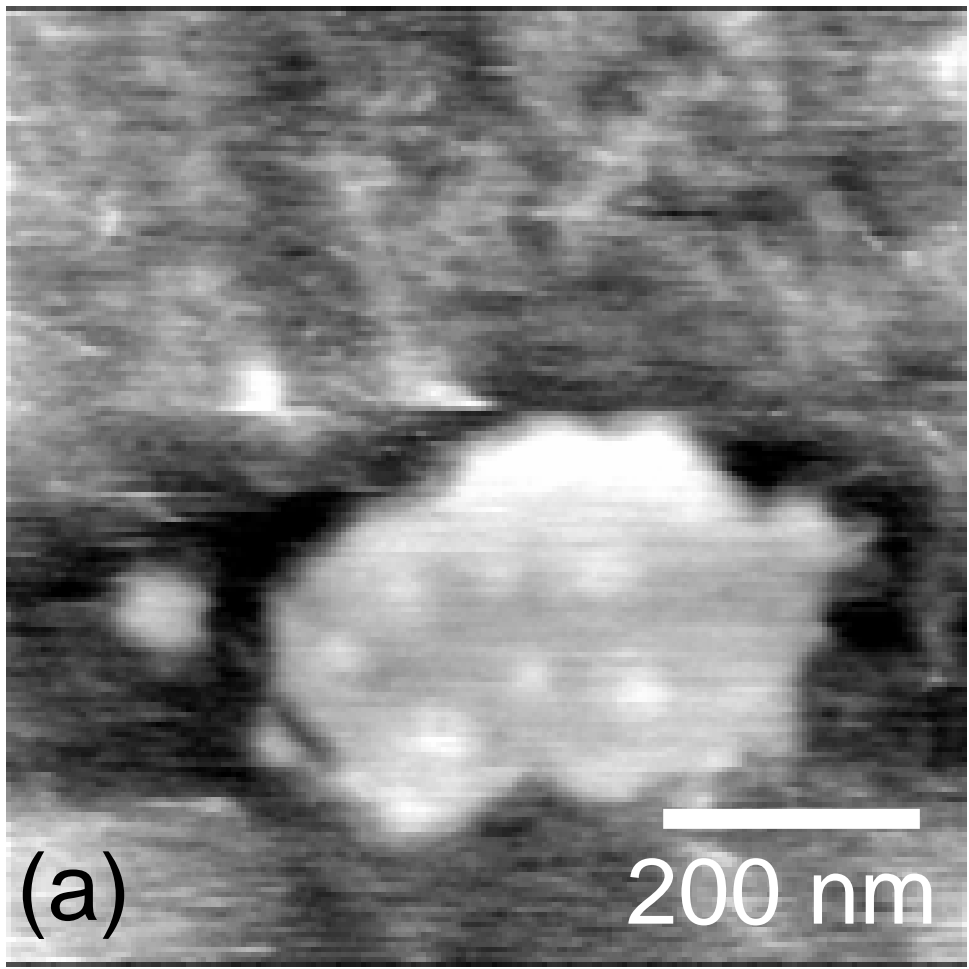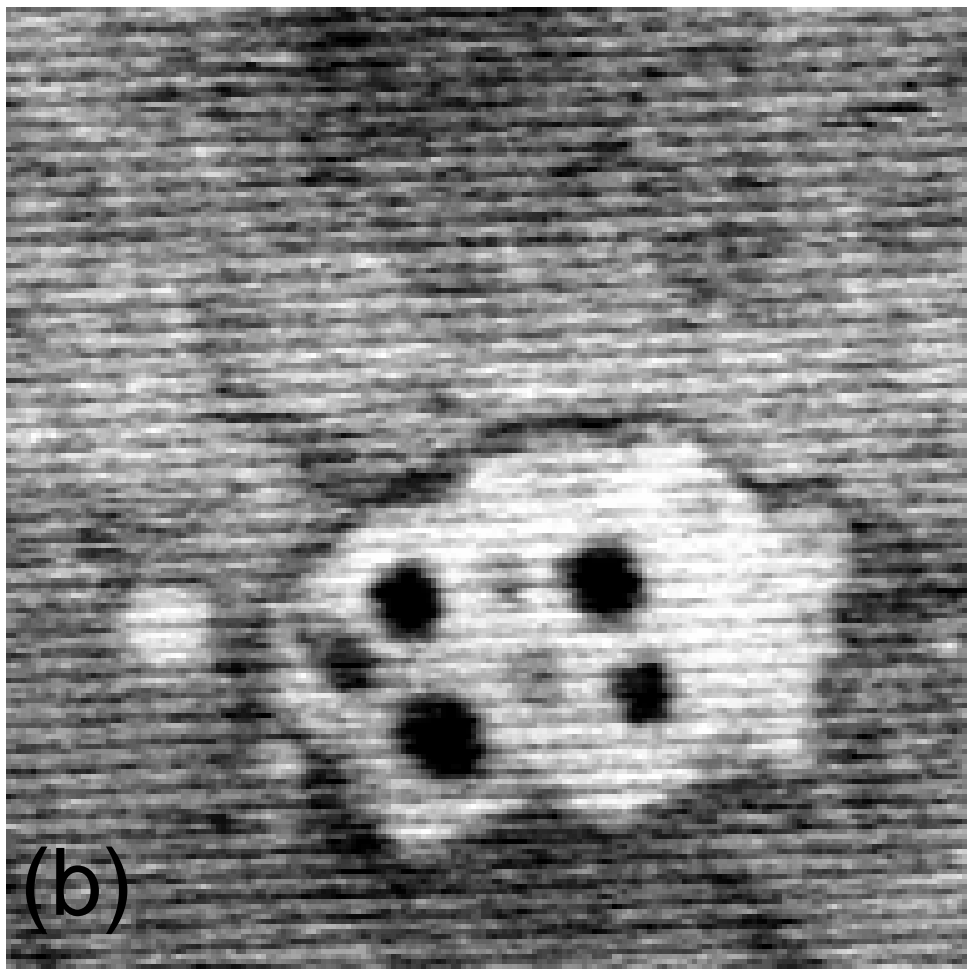

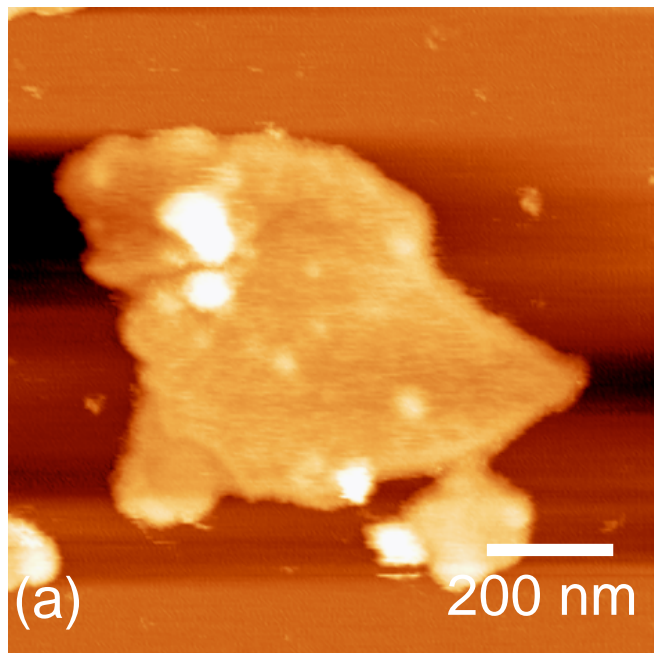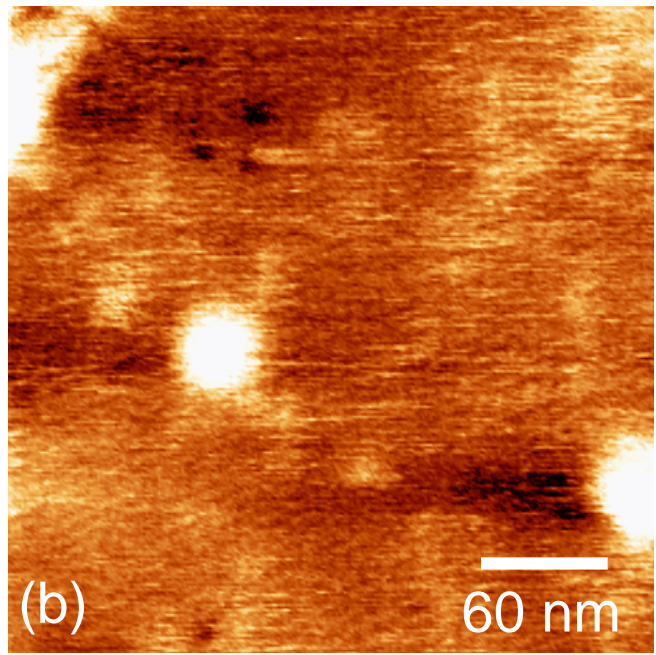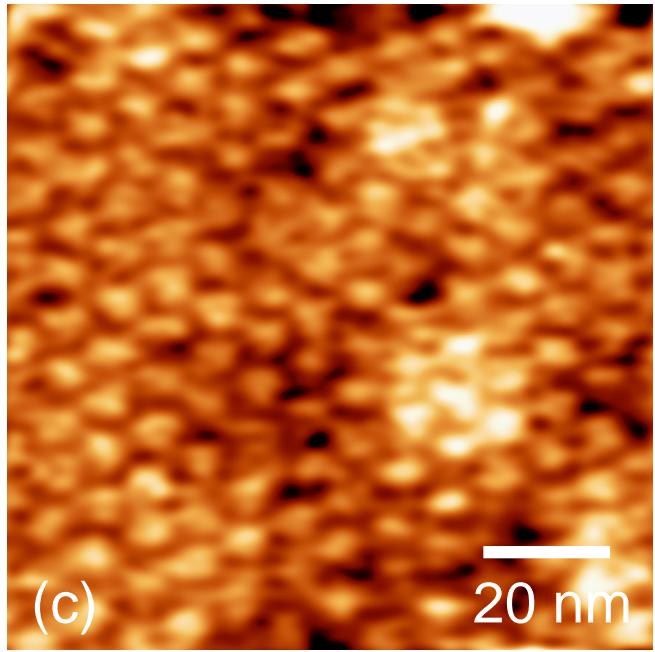

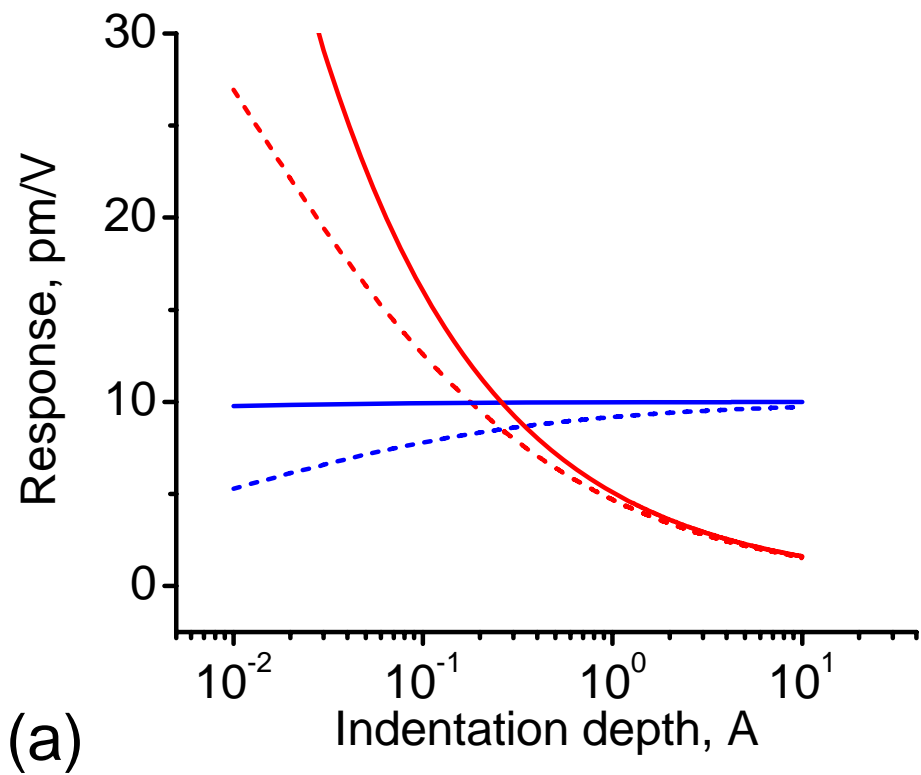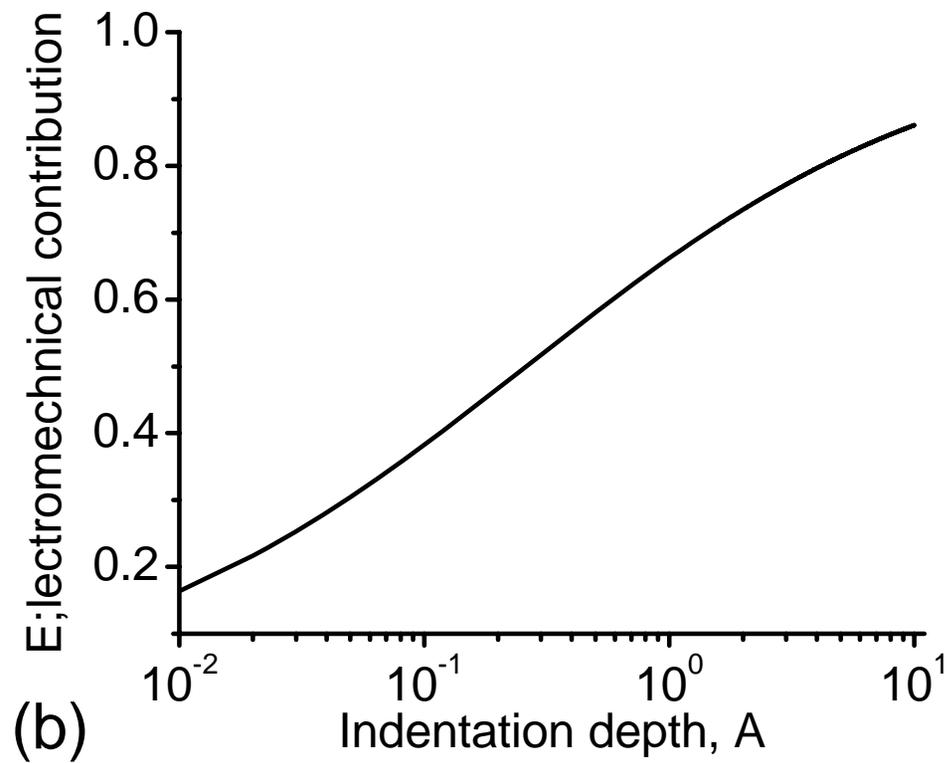

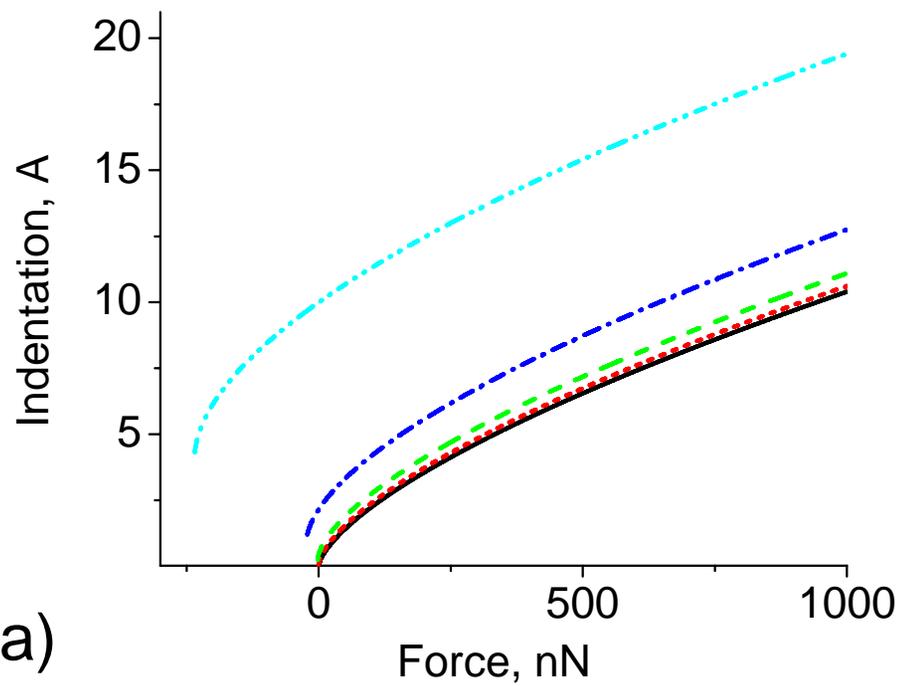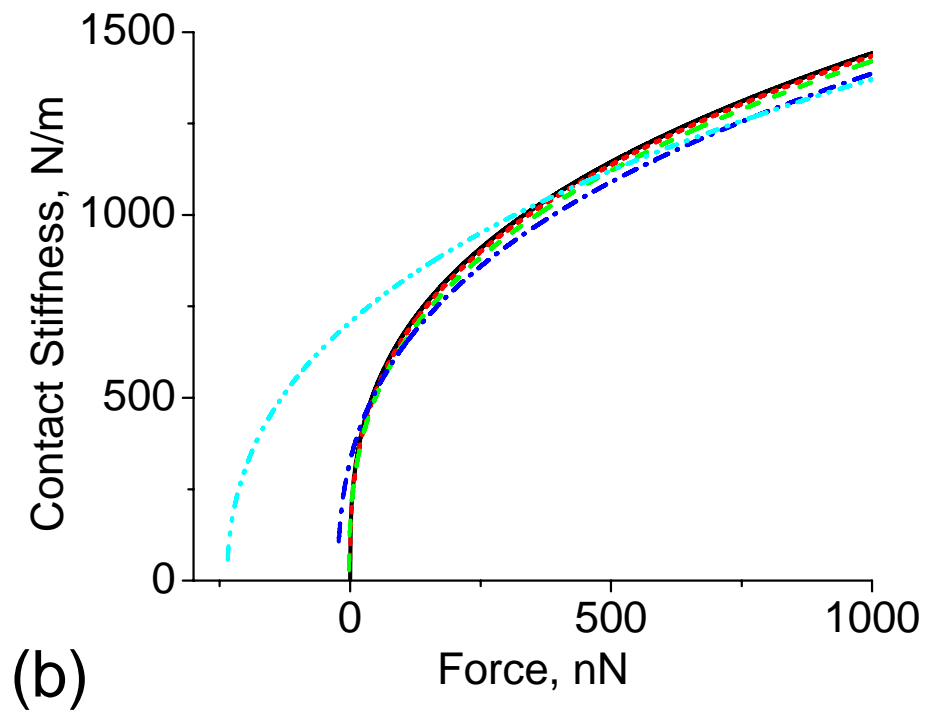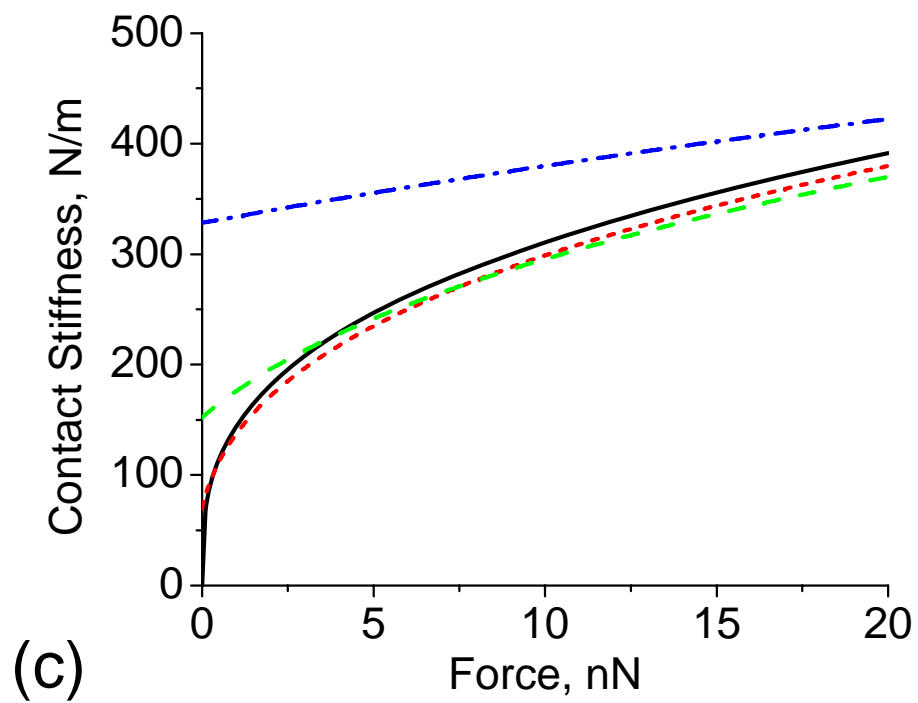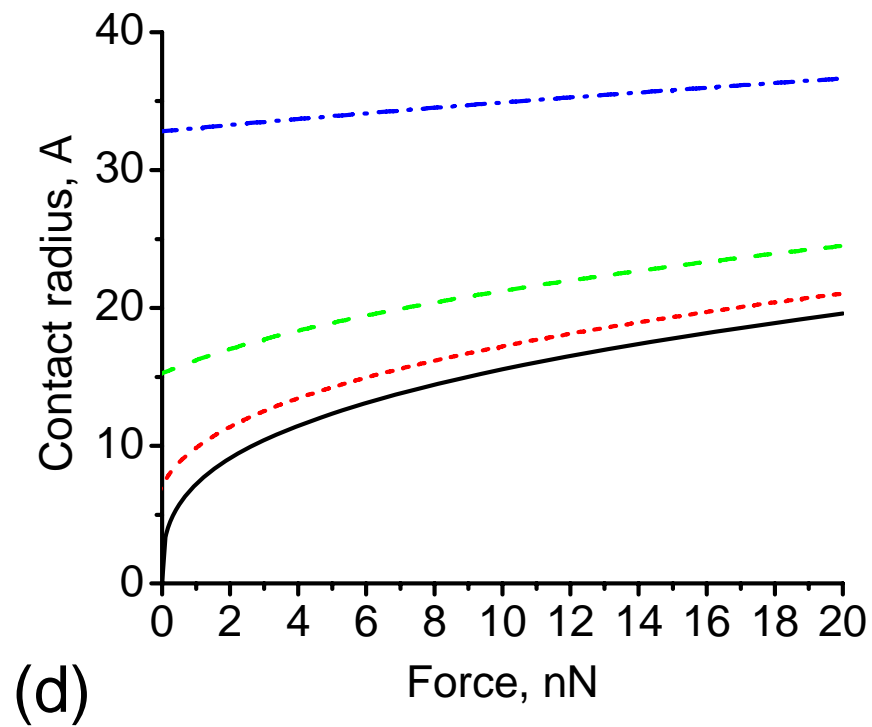